%% file: main.tex
\documentclass[sigconf]{acmart}
\acmConference[ISSTA 2024]{ACM SIGSOFT International Symposium on Software Testing and Analysis}{16-20 September, 2024}{Vienna, Austria}
\usepackage{booktabs}
\usepackage{subfigure}
\usepackage{stfloats}
\usepackage{listings}
\usepackage{listings-rust}
\usepackage{xcolor}
\usepackage[ruled, linesnumbered, commentsnumbered, noend]{algorithm2e}
\usepackage{amsthm}
\AtBeginDocument{%
  }

\setcopyright{acmcopyright}
\copyrightyear{2018}
\acmYear{2018}
\acmDOI{XXXXXXX.XXXXXXX}
\acmConference[arxiv 2023]{arxiv pre-print versions}{}{}

\acmPrice{15.00}
\acmISBN{978-1-4503-XXXX-X/18/06}

\begin{document}

\title{Fuzz Driver Synthesis for Rust Generic APIs}

\author{Yehong Zhang}
\affiliation{%
  \institution{School of Computer Science, Fudan University}
  \country{China}
}
\email{yehongzhang23@m.fudan.edu.cn}

\author{Jun Wu}
\affiliation{%
  \institution{School of Computer Science, Fudan University}
  \country{China}
}
\email{junwu@fudan.edu.cn}

\author{Hui Xu}
\authornote{Corresponding author.}
\affiliation{%
  \institution{School of Computer Science, Fudan University}
  \country{China}
}
\email{xuh@fudan.edu.cn}

\newtheoremstyle{definitionStyle} 
  {3pt} 
  {3pt} 
  {} 
  {} 
  {\bfseries} 
  {.} 
  {.5em} 
  {} 

\theoremstyle{definitionStyle}

\begin{abstract}

Fuzzing is a popular bug detection technique achieved by testing software executables with random inputs. This technique can also be extended to libraries by constructing executables that call library APIs, known as fuzz drivers. Automated fuzz driver synthesis has been an important research topic in recent years since it can facilitate the library fuzzing process. Nevertheless, existing approaches generally ignore generic APIs or simply treat them as normal APIs. As a result, they cannot generate effective fuzz drivers for generic APIs.

This paper studies the automated fuzz driver synthesis problem for Rust libraries with generic APIs. The problem is essential because Rust emphasizes security, and generic APIs are widely employed in Rust crates. Each generic API can have numerous monomorphic versions as long as the type constraints are satisfied. The critical challenge to this problem lies in prioritizing these monomorphic versions and providing valid inputs for them. To address the problem, we extend existing API-dependency graphs to support generic APIs. By solving such dependencies and type constraints, we can generate a collection of candidate monomorphic APIs. Further, we apply a similarity-based filter to prune redundant versions, particularly if multiple monomorphic APIs adopt the identical trait implementation. Experimental results with 29 popular open-source libraries show that our approach can achieve promising generic API coverage with a low rate of invalid fuzz drivers. Besides, we find 23 bugs previously unknown in these libraries, with 18 bugs related to generic APIs.

\end{abstract}




\maketitle

\section{Introduction}

Fuzzing is an automated technique for bug detection~\cite{mathis2020learning}. It tests a target executable program with random input and mutates the input according to the feedback information, such as code coverage. The technique can also be applied to library code by constructing executables that call the APIs of the library, known as fuzz drivers or fuzz targets. Due to its effectiveness and practicability, fuzzing has been widely employed by both industries~\cite{oss-fuzz, Honggfuzz} and open-source communities~\cite{AFL, LibFuzzer}.  

In this paper, we investigate library fuzzing for Rust libraries, with a specific focus on addressing the automated fuzz driver synthesis problem for generic APIs. The problem is important because Rust is a system programming language that places a strong emphasis on security and robustness. Bugs within Rust libraries have the potential to impact all downstream code. Generic APIs are widely employed in Rust libraries, offering developers a viable means to avoid repetitively implementing the same API for different types. A generic API generally has one or multiple parameters of generic types {\textit{e.g.,} \texttt{T}}, which will be monomorphized to concrete types upon invocation. However, existing work on fuzz driver synthesis (\textit{e.g.,} RULF~\cite{jiang2021rulf}) often neglects generic APIs or treats them merely as non-generic APIs. As a result, it either falls short in generating valid fuzz drivers for generic APIs or fails to produce diversified monomorphic versions for the same generic API.

A prevailing approach to automated fuzz driver synthesis involves traversing API-dependency graphs~\cite{jiang2021rulf}. However, when attempting to apply the method to generic APIs, two significant challenges emerge. Firstly, the type system of Rust is very powerful if not Turing-complete~\cite{Rust-TuringComplete}, leading to intricate dependencies for generic APIs and the difficulty of synthesizing valid API calls. In particular, one generic API may involve several type parameters (\textit{e.g.,} \texttt{T, U, G}), and each type parameter could be employed by several generic arguments (\textit{e.g.,} \texttt{a:T, b:Vec<T>}) of the function. To choose a concrete type (\textit{e.g.,} \texttt{i32}) for the generic type \texttt{T}, we should ensure that all these required concrete types (\textit{e.g.,} \texttt{a:i32, b:Vec<i32>}) can be produced by other APIs. Besides, these generic parameters should satisfy trait bounds, which may either bound the type parameters or generic parameters. Secondly, there could be numerous or even an infinite number of monomorphic APIs for a generic API. Fuzzing all these variations is impractical due to the time-intensive nature of the process. Note that it may take several hours or days to fuzz one monomorphic API. Therefore, we need to prioritize these monomorphic APIs and select a feasible number of APIs for fuzzing. To the best of our knowledge, no existing work has tackled these challenges.

This paper introduces a novel approach to automatically synthesize fuzz drivers for generic APIs. To tackle the first challenge, we carefully model the type constraint of generic types and extend the classic API dependency graph with generic APIs and types. By traversing the graph and solving type constraints, we can find all potential monomorphic APIs for creating fuzz drivers. In addressing the second challenge, we design a similarity-based pruning method to reduce the number of monomorphic APIs for fuzzing. Specifically, our method considers two concrete types as similar if they share the same trait implementation, indicating congruence in the behavior of the generic API.

We have implemented a prototype, named RULF+, building upon the state-of-the-art tool RULF, which lacks support for generic APIs. Our tool consists of 3K+ lines of Rust code, excluding the original RULF code. To assess the effectiveness of our approach, we conduct experiments with 29 open-source libraries, including 10 libraries randomly selected from Trophy-case\cite{trophy-case} and 19 other libraries randomly chosen from GitHub with 100+ stars and 5+ generic APIs. Experimental results show that our approach can achieve promising generic API coverage if the library does not depend on other third-party libraries. Furthermore, it generates a limited number of fuzz drivers, and most of the drivers are valid. Subsequently, we employ AFL++ to fuzz these drivers and find 23 bugs, including 18 bugs related to generic APIs. These results demonstrate that despite we have pruned numerous monomorphic APIs for efficiency, our approach is still able to discover bugs with carefully selected monomorphic APIs.

We summarize our contributions as follows.
\begin{itemize}
    \item This paper serves as the first attempt to study the fuzz driver synthesis problem for Rust generic APIs. We systematically model the type constraints and API dependencies of generic APIs. Moreover, we propose a novel algorithm to solve these constraints and dependencies in order to generate valid fuzz drivers. Experimental results show that our approach can achieve promising generic API coverage with a high rate of valid cases.
    \item We propose a novel similarity-based approach to prune redundant monomorphic APIs. Experimental results show that the remaining APIs are still able to discover bugs, although their numbers are limited.
    \item We have developed a tool, RULF+, and open-source released it to the community. 
\end{itemize}

\section{Problem}
\label{section:problem}
\subsection{Generic Type and Generic API}
Rust provides three types of generic parameters: type, constant, and lifetime\cite{generics}. However, in this paper, the term "generic" primarily refers to type generic parameters. This is because identifying type parameters is crucial for testing generic APIs. In Rust, type parameters can be used in the definitions of types, enums, and traits as substitutes for a specific type. 

Rust provides trait\cite{traits} to describe an abstract interface that types can implement. Trait can be regarded as a set of interface declarations. A type can implement a trait by implementing all interfaces declared in the trait. Based on the concept of trait, Rust provides a type constraint mechanism for the generic API called trait bounds\cite{Traitbounds}. A type parameter can specify the trait bounds to require the concrete type for this type parameter must implement the specific traits. If the type for the type parameter unsatisfies the requirement, the compiler will report a compilation error. On the other hand, as the compiler only accepts the type implementing these traits, the developer can safely invoke the interface of the traits for this type parameter in generic API. Trait bounds in Rust are composable, meaning if you want to specify that a type parameter \texttt{T} must implement both trait \texttt{A} and \texttt{B}, you can use the syntax \texttt{T:A + B} to express this requirement. Rust also provides a "where" clause, which allows programmers to declare trait bounds on any type or associated types\cite{generics}. This feature empowers Rust programmers to write type constraints flexibly and expressively.

Figure~\ref{fig:motivate} gives an example showing the usage of the generic API and the type in Rust. \texttt{map\_vec} is a generic API with two type parameters \texttt{T} and \texttt{U}. It receives an input with \texttt{Vec<T>} type, converts \texttt{T} to \texttt{U}, and returns \texttt{Vec<U>}. The API declares a trait bound \texttt{U: From<T>} to ensure that the type of \texttt{T} is capable of converting to the type of \texttt{U}. This trait bound also allows us to invoke the \texttt{U::from(item)} API to convert an item of type \texttt{T} to \texttt{U}. Line 8-9 shows the compiler accepts \texttt{map\_vec::<u8,i32>} and rejects \texttt{map\_vec::<i32,u8>}, since \texttt{i32} implements \texttt{From<u8>} but \texttt{u8} does not implement \texttt{From<i32>}.

\begin{figure}[ht]
    \centering
\begin{lstlisting}[language=Rust]
fn map_vec<T, U: From<T>>(vec: Vec<T>) -> Vec<U> {
    let mut ret=Vec::new();
    for item in vec.into_iter(){
        ret.push(U::from(item));
    }
    ret
}
let _ = map_vec::<u8,i32>(vec![0u8]);  // accept
let _ = map_vec::<i32,u8>(vec![0i32]); // compile fail
\end{lstlisting}
    \caption{Example of generic usage in Rust.}
    \label{fig:motivate}
\end{figure}

The paradigm of generic programming is immensely powerful, yet it introduces a new security issue during the process of software development. To support generic programming, compilers implicitly generate multiple monomorphic APIs for a generic API based on the needed type in the process of monomorphization. An elusive bug can hide on a certain monomorphic API and cannot be detected in other monomorphic APIs. Due to the transparency of the monomorphization process to programmers, it is difficult for them to discover and test such bugs.

\subsection{Fuzz Driver Synthesis with Generic API}
\subsubsection{Generic API Dependency Graph}
We extend the API dependency graph to a generic API dependency graph so that we can model all APIs, including generic APIs and dependencies among them. 

\begin{definition}[generic API dependency graph]
A generic API dependency graph is defined as a directed graph $\mathcal{G}=(N_A, N_T, E)$ that:

\begin{align*}
N_A & =A\cup \mathcal{A} \\
N_T & =T \cup \mathcal{T} \\
E & =C\cup P\cup M
\end{align*}

where:
\begin{itemize}
    \item $A$ is the set of non-generic API nodes.
    \item $\mathcal{A}$ is the set of generic API nodes.
    \item $T$ is the set of non-generic type nodes.
    \item $\mathcal{T}$ is the set of generic type nodes.
    \item $C$ is the set of consumer edges, where $C\subseteq N_A \times N_T$.
    \item $P$ is the set of producer edges, where $P\subseteq N_T \times N_A$.
    \item $M$ is the set of match edges, where $M\subseteq T\times \mathcal{T}$.
\end{itemize}
\end{definition}

The definition of $A$, $\mathcal{A}$, $T$, $\mathcal{T}$ is obvious. We distinguish three types of edges in the generic API dependency graph. The consumer edge $t\rightarrow a$ represents API $a$ requires an instance of type $t$. The producer edge $a\rightarrow t$ represents API $a$ produce(or return) an instance of type $t$. Note that for consumer and producer edges, if $a\in A$, then we have $t\in T$, since the non-generic API cannot consume or produce a generic type. The match edge $t_1\rightarrow t_2$ represents a concrete type $t_1$ that can match the generic type $t_2$, that is, type $t_1$ may serve as an instance of type $t2$. For example, \texttt{Box<i32>} can match \texttt{Box<T>}, but \texttt{Vec<i32>} cannot match \texttt{Box<T>}. The match edges serve as a bridge between the concrete type and the generic type, and it is crucial when we try to infer the generic API dependency.

\begin{figure}[ht]
    \centering
\begin{lstlisting}[language=Rust]
struct Ty1;
struct Ty2;
trait A{}
trait B{}
impl A for Ty1 {}
impl A for Ty2 {}
impl B for Ty2 {}
fn f1() -> Ty1;
fn f2() -> Ty2;
fn f3<T:A>(a1:T)->T;
fn f4<T> (a1:T)->Vec<T>;
fn f5<T:A+B> (a1:Vec<T>);
\end{lstlisting}
    \caption{Source code for an example library.}
    \label{fig:api-graph-src}
\end{figure}

\begin{figure}[ht]{
    \centering
    \includegraphics[width=0.8\linewidth]{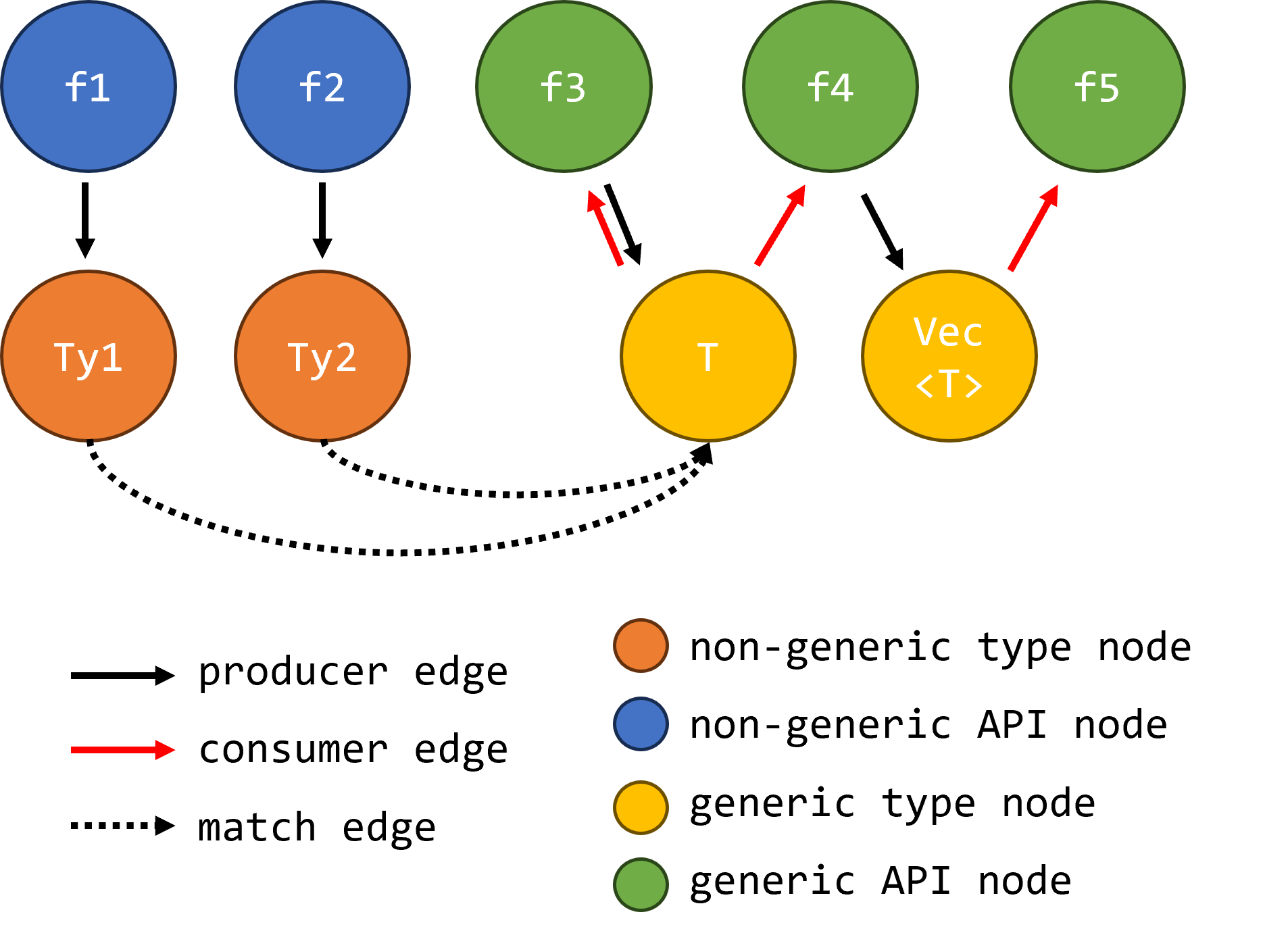}
    \caption{Generic API dependency graph for figure~\ref{fig:api-graph-src}.}
    \label{fig:generic-api-graph}
}
\end{figure}

Figure~\ref{fig:generic-api-graph} demonstrates a generic API dependency graph for an example library. The library source code is shown in figure~\ref{fig:api-graph-src}. The graph contains 2 non-generic nodes and 3 generic nodes. There are 2 concrete types \texttt{Ty1}, \texttt{Ty2} and 2 generic types \texttt{T} and \texttt{Vec<T>}. The consumer and producer edges are constructed by the input types and return types for each API. Finally, there are two match edges(\texttt{Ty1}$\rightarrow$ \texttt{T}, \texttt{Ty2}$\rightarrow$ \texttt{T}) that represent both \texttt{Ty1} and \texttt{Ty2} can match with generic type \texttt{T}.

\subsubsection{The Challenges of Fuzz Driver Synthesis}
\label{sec:fuzz-driver-synthesis}
A fuzz driver for a library can be viewed as a sequence of library API calls. Thus, we model the task of fuzz driver synthesis as the task of API sequence generation.

\begin{definition}[API Sequence]
An API sequence is a list of APIs. For example, $S=[api_1,api_2,\cdots,api_n]$ is an API sequence.
\end{definition}
 
\begin{definition}[Reachable API]
\label{def:reachable-api}
An API in the API sequence is said to be reachable if each of its input satisfy any one of the following criteria:
\begin{itemize}
    \item The type of this input is primitive.
    \item The type of input is non-generic, and there exists a producer edge from any previous API in the API sequence to this input type.
    \item The type of input is generic, and there exists a concrete type $t$ that exists a producer edge from any previous API in the API sequence to type $t$ and exists a match edge from $t$ to this input type.
\end{itemize}
\end{definition}

\begin{definition}[Valid API Sequence]
An API sequence is said to be valid if all APIs in this sequence are reachable.
\end{definition}

To synthesize valid fuzz drivers, we have to generate corresponding valid API sequences. A valid API sequence ensures that all APIs can obtain instances of their input types from previous APIs or the fuzzer so that the fuzz driver can be compiled successfully. However, the generic APIs in a valid API sequence should be replaced with specific monomorphic APIs so that we can determine the exact data dependencies. In other words, we should determine the concrete types for every type parameter in the generic API and ensure the type combination for type parameters can satisfy the trait bounds of the generic API.

This process contains several challenges to the API sequence generation that no previous work has tackled. We summarize two fundamental challenges as follows:


\paragraph{\textbf{Selecting monomorphic APIs for generic APIs}} The type system of Rust is very powerful, leading to the difficulties of selecting a reachable and valid monomorphic API for a generic API. In particular, the generic API dependency graph is not steadfast but keeps changing during the monomorphization process. The number of type nodes \textit{t} would continuously grow because there could be a number of new types generated if the return type of a generic API is generic. For example, in the example library of Figure~\ref{fig:generic-api-graph}, calling \texttt{f1},\texttt{f4} sequentially can produce type \texttt{Vec<Ty1>}, which have not appeared in the graph before. This constructs two new match edges \texttt{Vec<Ty1>}$\rightarrow$\texttt{Vec<T>} and  \texttt{Vec<Ty1>}$\rightarrow$\texttt{T}. 

Furthermore, there could be complicated trait bounds. The trait bounds can either be applied to type parameters (\textit{e.g.,} \texttt{T:Bound1}) or generic parameter types (\textit{e.g.,} \texttt{where Vec<T>:Bound1}). One generic API may have multiple generic parameters and trait bounds, and these bounds could be related to each other. Searching valid concrete types based on the generic API dependency graph and satisfying all bounds is not easy.


\paragraph{\textbf{Monomorphic API explosion}} The number of valid monomorphic APIs can be large or even infinite, making it impractical to cover all monomorphic APIs in fuzzing. Supposing a generic API has $m$ generic parameters, and each type parameters has $n$ valid monomorphic versions, the number of valid monomorphic versions for the API is exponential to $m$, \textit{i.e.,} $n^m$. Besides, a generic API returning a generic type can produce an infinite number of new types. Taking Figure~\ref{fig:generic-api-graph} as an instance, repeatedly passing the return type of \texttt{f4} as an input to itself, we can produce type \texttt{Vec<Ty1>}, \texttt{Vec<Vec<Ty1>}\texttt{>}, \textit{etc}.

\section{Approach}

\begin{figure*}[htbp]
\small
    \centering
    \includegraphics[width=0.9\linewidth]{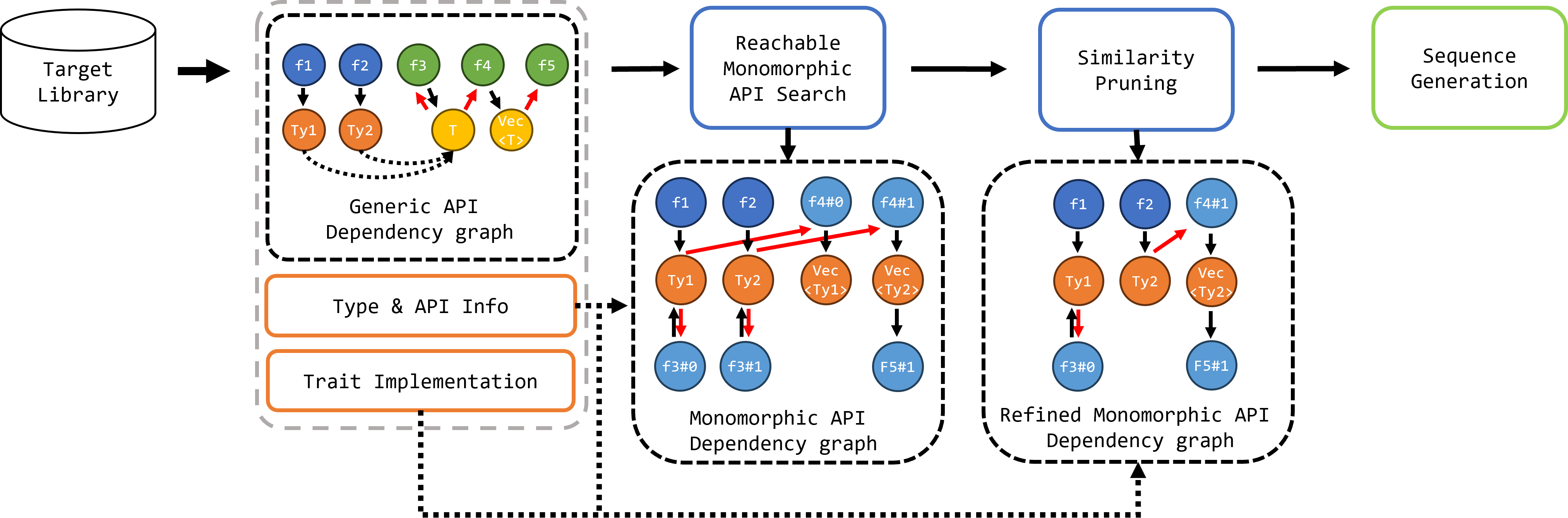}
    \caption{Overview of our approach.}
    \label{fig:approach}
\end{figure*}


\subsection{Overview}
Figure~\ref{fig:approach} overviews our approach. We propose a two-stage approach to tackle two challenges. To tackle the first challenge, we apply the reachable monomorphic API search to resolve all valid and reachable monomorphic APIs. Our reachable monomorphic API search infers the possible monomorphic APIs from the match edge from the reachable concrete type to the generic arguments of the generic API. This is crucial to our approach since it ensures the reachability of generated monomorphic APIs and avoids enumerations of all possible type combinations. We also apply the trait bounds check in the reachable monomorphic API search to ensure all generated monomorphic APIs are valid. After searching, the generic API dependency graph is transformed into a monomorphic API dependency graph. To tackle the second challenge, we apply similarity pruning to reduce the number of monomorphic APIs. We identify similar behaviors among monomorphic APIs. If two monomorphic APIs share the same implementation, we consider them similar. Based on this assumption, we prune monomorphic APIs to reduce redundant testing of similar behaviors. We obtain a refined monomorphic API dependency graph after this stage. Finally, we do API sequence generation and fuzz driver synthesis based on RULF. We introduce this stage in the next section.

\subsection{Reachable Monomorphic API Search}
We start by defining several concepts to introduce our algorithm.


\begin{definition}[Monomorphic solution]
    A \textit{monomorphic solution} of a generic API refers to a specific substitution scheme for all type parameters in the generic API. Formally, we present a monomorphic solution as a tuple $(t_1,t_2,\dots)$, where $t_i$ represents the concrete type to replace the i-th type parameter. The type parameter is ordered by the declared order. 
\end{definition}
\paragraph{Example} A monomorphic solution for API \texttt{foo<T,U>(T)->U} can be \texttt{(f32,i32)}, which means to substitute \texttt{T} with concrete type \texttt{f32} and \texttt{U} with \texttt{i32}. This monomorphic solution also corresponds to a monomorphic API \texttt{foo(f32)->i32}.

\begin{definition}[Top and bottom]
We use $\top$ to represent an arbitrary type for a type parameter, and $\bot$ to represent a bottom type for a type parameter. 
\end{definition}
\paragraph{Example} For function \texttt{foo<T,U>(T,U)}, \texttt{($\top$,$\top$)} represents arbitrary type combination for type parameters is valid. For function \texttt{foo<T:A,U:A>(T,U)}, \texttt{($\bot$,$\bot$)} represents no type combination for parameters is valid. The merge and union operation for the monomorphic solution set with $\top$ is also similar, by treating $\top$ as an arbitrary type that can match any type. For instance, $\{(\top,i32)\}\cap\{(f32,i32)\}=\{(f32,i32)\}$, $\{(\top,i32)\}\cup\{(f32,i32)\}=\{(\top,i32)\}$. For $\bot$, we regard $\bot$ can not matching with any type. For instance, $\{(\bot,\bot)\}\cap\{(i32,f32),(f8,i8)\}=\{(i32,f32),(f8,i8)\}$.


\begin{definition}[Match function]
we define function \texttt{Match($t_1$,$t_2$)}, where $t_1\in T$, $t_2\in \mathcal{T}$ and $t$ is a generic argument for a generic API $gapi$. If $t_1\rightarrow t_2\notin M$, then the match function returns $(\bot,\bot)$. If $t_1\rightarrow t_2\in M$, the match function returns a possible monomorphic solution for $gapi$ regarding $t_1$ as the concrete type of $t_2$. 
\end{definition}
\paragraph{Example} for function \texttt{foo<T,U>(Vec<T>,U)}, \texttt{Match(Vec<i32>,}

\texttt{Vec<T>)} returns \texttt{(i32,}$\top$\texttt{)}, and \texttt{Match(Box<i32>,Vec<T>)} returns \texttt{($\bot$,$\bot$)}, as \texttt{Vec<T>} and \texttt{Box<i32>} has different outer type.

%

\subsubsection{Algorithm}
\input{pseudo/reachability-v2}

To meet the requirement of validity and reachability, We propose an iterative algorithm for obtaining reachable monomorphic solution sets for all generic APIs by utilizing the reachable information from type dependencies. our algorithm first searches the set of monomorphic solutions for each generic argument for generic APIs. Then our algorithm merges all sets of monomorphic solutions for a generic API, getting the largest common monomorphic solutions set. The monomorphic solutions in this set ensure the reachability of each generic input. We then check trait bounds to ensure the validity of monomorphic solutions. We discuss the details of the algorithm below.

Algorithm~\ref{alg:reachable-search} introduces our algorithm. We define $reachableType$ as the set of reachable types that our algorithm has found. Once we find a new reachable API (non-generic API or monomorphic API), we add the return type of the API to $reachableType$. The main loop (line 2-21) continues to run until $reachableType$ remains unchanged. In each iteration, the algorithm tries to find new reachable APIs for non-generic API and generic API, respectively. First, the algorithm checks all non-generic APIs whether it is reachable now. If an API is reachable, then we add the return type of this API to the $reachableTypes$ (line 3-5). %

Then the algorithm checks whether a new monomorphic solution for a generic API is reachable. We check each argument for each generic API. If the argument is generic, we perform \texttt{Match} between this generic argument and all types in $reachableTypes$ and we union the returning set from \texttt{Match} (line 8-13) to a larger set. Since type in $reachableType$ can be instantiated in some API sequences, for all monomorphic solutions in this set, it must be reachable in some API sequences. However, some monomorphic solutions may only be reachable for partial generic arguments in a generic API. Consequently, we merge the monomorphic solution sets for each generic API, yielding the largest common monomorphic solution set (line 13). 

\begin{figure}[ht]
    \centering
    \includegraphics[width=0.85\linewidth]{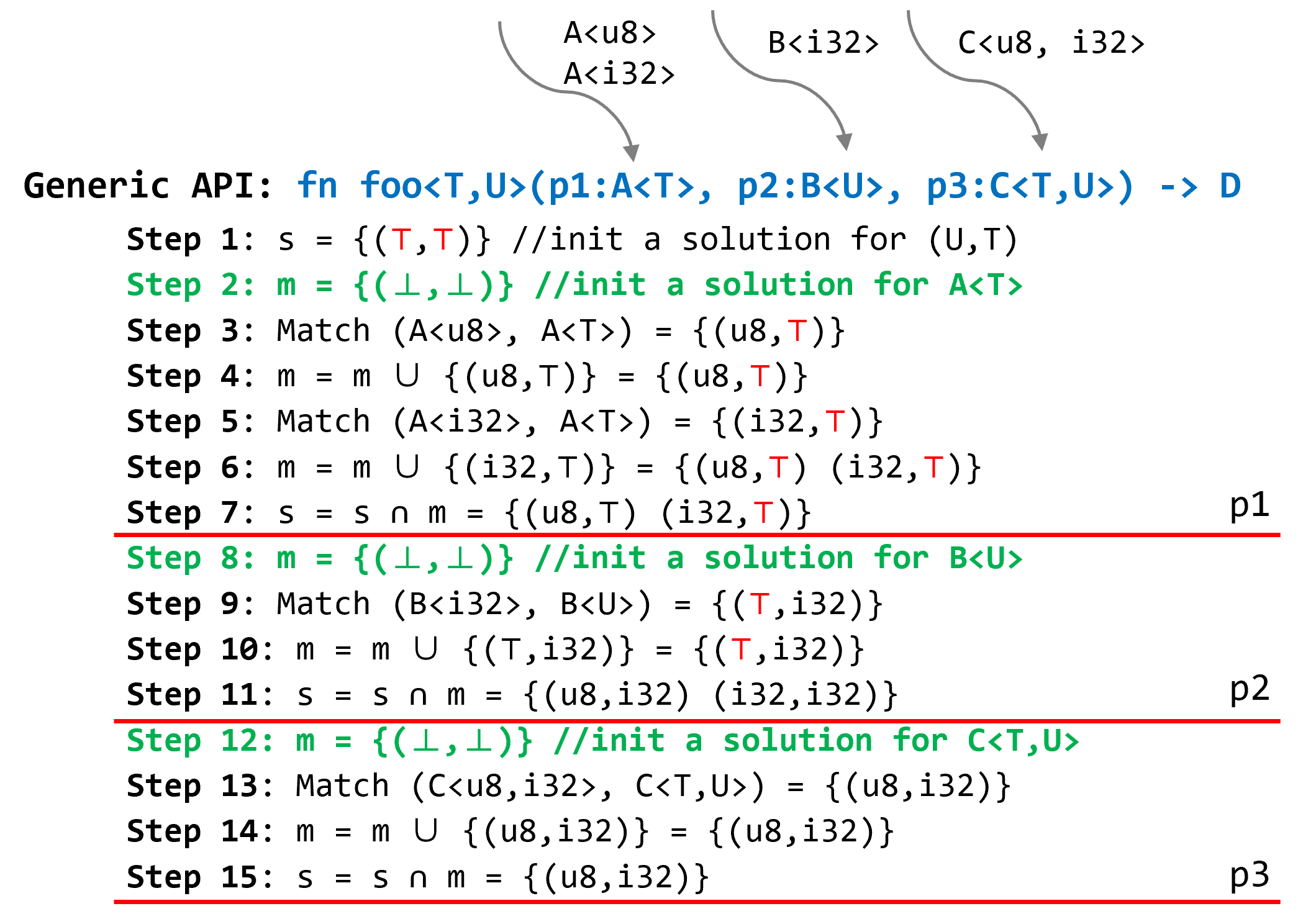}
    \caption{Example of applying algorithm~\ref{alg:reachable-search}. All $\top$ notions are marked red.}
    \label{fig:case_solvegeneric}
\end{figure}

Figure~\ref{fig:case_solvegeneric} offers an example to better illustrate the union and merge processes of monomorphic solution sets. Step 1 initiates the largest common monomorphic set $s$. Steps 2-7 perform \texttt{match} for $p1$ and all types in $reachableTypes$ and perform a union operation. Steps 8-11 and step 12-15 repeat the same logic as steps 2-7 for $p2$ and $p3$, respectively. In step 15 we obtain the final largest common monomorphic solution set $s=\{$\texttt{(u8,i32)}$\}$, meaning only \texttt{(u8,i32)} is a valid and reachable monomorphic API for the generic API.

To ensure the validity of the monomorphic solutions, we employ a trait bounds check for each monomorphic solution (line 15). Note that we still need to check the reachability for each monomorphic API, since the non-generic argument in the API may be unreachable. To ensure the algorithm will not generate infinite monomorphic solutions for some generic APIs, we set a maximum threshold $maxDepth$ for type nested and prohibit generating the monomorphic solution containing types that exceed this threshold. It is reasonable for generating fuzz drivers, as overly complex monomorphic APIs typically do not increase code coverage.

\subsection{Similarity Pruning}

\label{sec:pruning-movtivate}


Intuitively, the total number of codes(or implementations) is limited, so that there may be some monomorphic APIs share the same implementation. Thus, their behavior may be similar in a generic API with certain trait bounds. We analyzed real-world libraries and identified two common practices that lead to similar behaviors: blanket implementation and default implementation. Figure~\ref{fig:shared-behavior} is an example of how these common practices create similar behaviors. Trait \texttt{Foo} provides a default implementation for method \texttt{foo}. When type \texttt{A} and \texttt{B} implement \texttt{Foo}, they do not provide a specific implementation for \texttt{foo}, resulting in sharing the same default implementation. On the other hand, the example employs a blanket implementation that provides the implementation of \texttt{Bar} for both \texttt{A} and \texttt{B}, resulting in similar behaviors. 

\begin{figure}[ht]
\begin{lstlisting}[language=Rust]
struct A;
struct B;
trait Foo{
    fn foo()->&'static str { "default impl for Foo" }
}
trait Bar{
    fn bar()->&'static str;
}
// Situation 1: similar behavior by default implementation
impl Foo for A {}
impl Foo for B {} 
// Situation 2: similar behavior by blanket implementation
impl<T:Foo> Bar for T{
    fn bar() -> &'static str { "blanket impl for Bar" } 
}
\end{lstlisting}
    \caption{Example of how default and blanket implementation create similar behaviors.}
    \label{fig:shared-behavior}
\end{figure}

In the context of fuzzing, repeatedly testing similar behaviors can lead to low efficiency. Therefore, we propose a novel pruning algorithm to remove similar monomorphic APIs, reducing the number of generated drivers and improving fuzzing efficiency. 


\subsubsection{Algorithm}

The idea of our algorithm is to cover all implementations with the minimum number of monomorphic solutions, since testing the same implementation across multiple fuzz drivers may be inefficient. The algorithm initially treats all APIs as non-preserved APIs, subsequently selecting the minimum number of monomorphic APIs for each generic API to cover all implementations under the generic API trait bounds. To maintain the reachability of reserved monomorphic APIs, we also reserve APIs that can produce instances for the argument of reserved APIs, until all arguments of reserved APIs are reachable.

\begin{definition}[Trait implementation set]
We define $Impls(gapi,m)$ as the set of trait implementations of the monomorphic solution $m$ for the generic API $gapi$. A trait implementation $i\in Impls(gapi,m)$ if the corresponding trait is declared within the trait bounds of the generic API.
\end{definition}
For example, consider API \texttt{fn foo<T:Tr1,U>(T,U)}, and $m_1=(A,B), m_2=(C,D)$. The $Impls(gapi,m_1)=\{$ A's implementation of Tr1$\}$, while $Impls(gapi,m_2)=\emptyset$. If two monomorphic solutions share the same implementation $i$, then $i\in Impls(gapi,m_1)\wedge i\in Impls(gapi,m_2)$.

\input{pseudo/pruning}

Algorithm~\ref{alg:similarity-pruning} introduces our algorithm. The algorithm starts with an empty reserved API set $R$.

The algorithm first performs \texttt{GetMinimalSetCover} to initialize $R$ for each generic API $gapi$ (line 2-3). \texttt{GetMinimalSetCover} receives an generic API $gapi$, select a set cover $S$ from $MO(gapi)$ and add $\{MonoAPI(mo)| mo\in S\}$ to the $R$. $S$ is a set cover that the union of $Impl(m,gapi)$ for all $m\in S$ is equal to the universal set of $m\in MO(gapi)$. 

To reduce the size of the initial $R$, we wish to select the minimal subset of monomorphic solutions that cover a universal set of implementations. This is known as the \textbf{set cover problem (SCP)} that is NP-hard\cite{SCP-complete}, so we use a simple greedy algorithm to select monomorphic solutions for $gapi$ (line 14-17). In each turn, we select $mo$ with maximum uncovered implementations until all implementations are covered.

As we initialize all APIs as non-reserved, it is possible that the arguments of these APIs cannot be produced by $api\in R$ temporarily. We conduct an iterative search across all APIs to propagate the reserved types through the backward of producer edges (line 4-11). We add all arguments of initial reserved APIs to a set $requireType$, representing the types that we need to find other APIs to produce an instance for (line 21-22). To prevent an excessive number of reserved APIs, we add types that have already found the producer to the $produceTypes$. During each iteration, if an API or a monomorphic API can produce a type $ty$ that $ty\in requireTypes\wedge type\notin produceTypes$, we add this API to the set of reserved APIs $R$ (line 23-31). The iteration runs until $reserveType$ and $produceType$ remain unchanged.

\section{Implementation}
\label{sec:impl}
We implement a prototype called RULF+, since we implement our approach based on the codebase of RULF. RULF is a tool that automatically generates fuzz drivers for the Rust library. It leverages rustdoc\cite{Rustdoc} to extract all API information from the library, then construct an API dependency graph and generate fuzz drivers. However, RULF only extracts non-generic APIs and generates fuzz drivers for them, without any consideration of generic APIs. We integrate our approach with RULF by transforming the generic API dependency graph into the refined monomorphic API dependency graph so that RULF can work without any type parameter. Besides, we update the compiler version used by RULF to 1.66.0-dev, enabling our tool to analyze the latest versions of libraries. We write about 3K+ lines of Rust code to implement our approach.

\begin{figure}[ht]
\small
    \centering
    \includegraphics[width=0.9\linewidth]{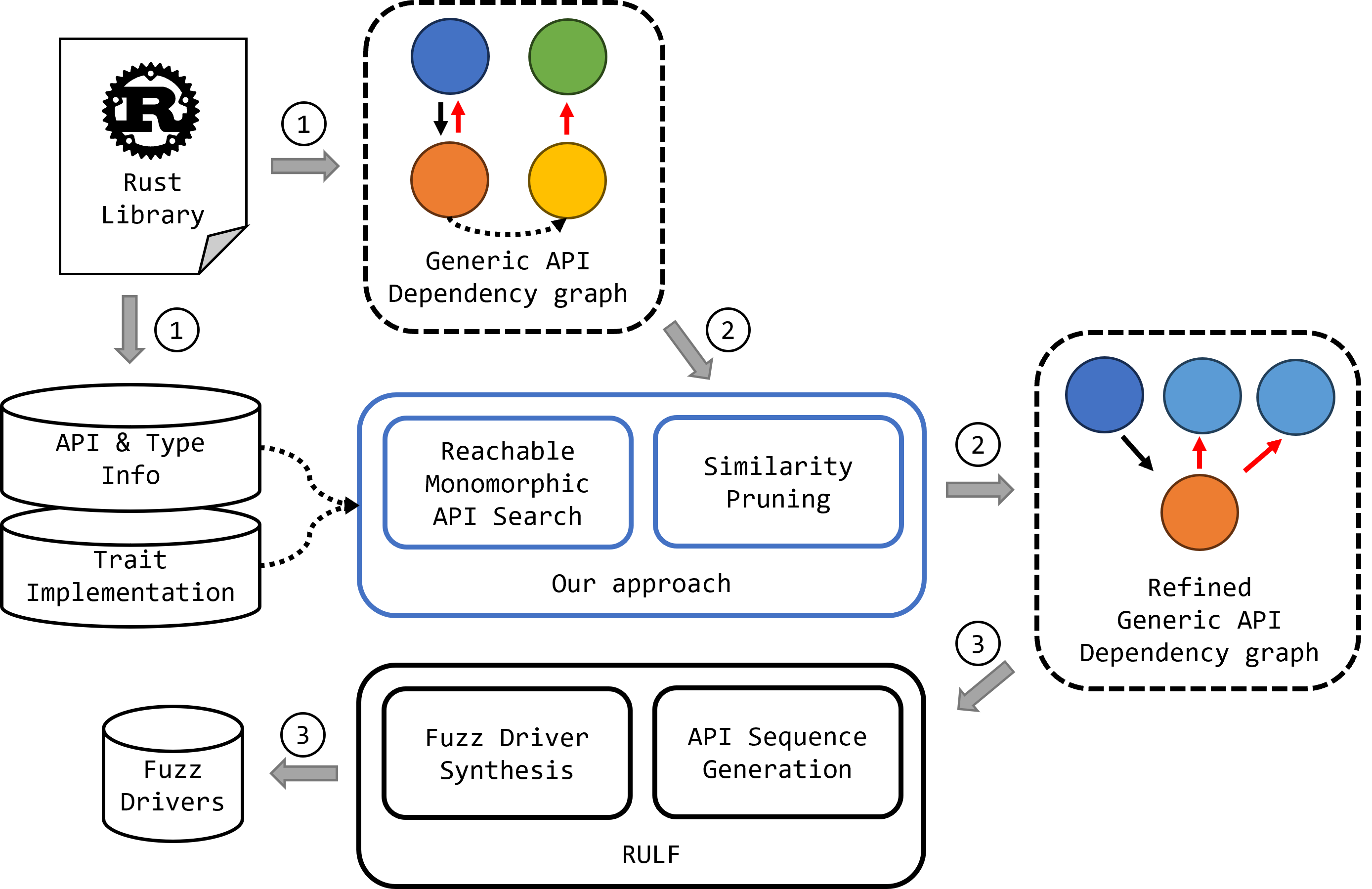}
    \caption{Workflow of our implementation.}
    \label{fig:workflow}
\end{figure}

The figure~\ref{fig:workflow} overviews the workflow of our implementation. Our workflow contains the following steps:

\begin{enumerate}
    \item \textbf{Library information extraction.} We extract API signatures, type specifications, and trait implementations from the target library.
    \item \textbf{API graph monomorphization.} We perform reachable monomorphic API search and similarity pruning to transform the generic API dependency graph into a refined generic API dependency graph in this stage.
    \item \textbf{API Sequence Generation and fuzz driver synthesis.} We integrate our approach with RULF, feeding the refined monomorphic API dependency graph to generate API sequences and synthesize the fuzz drivers. 
\end{enumerate}

We discuss the details of each step in our implementation in the rest of the section.

\subsection{Library Information Extraction}
We collect and process several types of information from the target library. For non-generic API, we simply collect the API signature. For generic API, we collect the generic API signature, with the type parameter declaration and trait bounds declaration. While collecting API information, we regard the implementation of the same trait for various types as separate APIs. For instance, if \texttt{A} and \texttt{B} both implement \texttt{foo} in trait \texttt{Bar}, we regard \texttt{A::foo} and \texttt{B::foo} as two unique APIs. Moreover, we excluded certain common implementations of standard library(std) traits which used for common purposes, e.g. \texttt{Debug} and \texttt{Display} traits, as these traits are typically unrelated to the primary objectives of the library. 

The trait implementation information we collect can be represented as a tuple $(type, trait, impl\_id)$, meaning that $type$ implements $trait$ with $impl\_id$. $impl\_id$ is a unique id we extract from rustdoc. If two types share the same implementation as we mention in~\ref{sec:pruning-movtivate}, they will get the same $impl\_id$.

As the limitation of rustdoc, we can only gather information that should be documented. As a result, the third-party API information, including std API information can not be obtained by our tool. However, some std APIs are widely used in public libraries, e.g. \texttt{std::Vec} and \texttt{std::String}. To alleviate this situation, we employed the re-exports technique\cite{re-exports} in Rust, inserting a small code snippet to allow us to obtain the std API information. Figure~\ref{fig:re-exports} shows the code snippet we insert into the target library. During our experiments, we observed that employing this technique substantially enhances API coverage in certain libraries.

\begin{figure}[ht]
\begin{lstlisting}[language=Rust]
pub use std::vec::Vec;
pub use std::string::String;
pub use std::collections::hash_map::DefaultHasher;
pub use std::io::Write;
pub use std::io::Read;
\end{lstlisting}
    \caption{Re-exports code snippets.}
    \label{fig:re-exports}
\end{figure}

\subsection{API Graph Monomorphization}
\label{sec:api-graph-monomorphization}
Once we collect all the needed information from the target library, we start transforming the generic API dependency graph into a monomorphic API dependency graph by applying reachable monomorphic API search and similarity pruning. The approach has been fully covered in the previous section, so we only discuss several details in this section.

To apply the reachable monomorphic API search, we need to match types for each generic argument with current reachable types. We predefined a set of transformation rules to increase the success rate of this process. We first analyze whether $ty$ can be matched with $arg$ by applying one or multiple predefined transformation rules. If so, we first apply these transformation rules and then perform $match(arg,ty)$. We list all predefined transformation rules in table~\ref{tab:transformation}. 

\input{tables/transformation}

To check the trait bounds, We implement a trait bounds checker in our tool. However, due to the limitation of rustdoc, we can only obtain a subset of the trait implementation and part of the associated item\cite{associate-items} information in traits. As we intend to reach a high generic API coverage as much as possible, the trait bounds checker we implement is unsound, meaning that it may produce false positive results in some cases. However, it can cover more APIs in most of the cases. The experimental results also demonstrate that the number of generated invalid fuzz drivers is actually quite low in real-world libraries.

\subsection{API Sequence Generation}
\label{sec:api-sequence-generation}
After we transform the generic API dependency graph into a refined monomorphic API dependency graph, we utilize RULF to generate API sequences and synthesize the fuzz drivers. The sequence generation of RULF ensures that each API will be covered in fuzz drivers at least once, as long as there exist valid API sequences to invoke the API. Since our method ensures that all monomorphic APIs are reachable, each one can be ultimately covered by the API sequences.

To prevent an excessive number of API sequences, we limit the number of API sequences to 300. To test generic functions, we impose a constraint that the API sequence must include at least one monomorphic API. By removing this constraint, our tool can generate API sequences for all possible APIs in the library.

\section{Evaluation}
\input{tables/result}




We evaluate our approach with three research questions:
\begin{itemize}
    \item \textbf{RQ1}: Can our approach generate valid fuzz drivers for generic APIs?
    \item \textbf{RQ2}: Does our pruning algorithm improve efficiency? 
    \item \textbf{RQ3}: What kinds of bugs our approach can find in the real-world libraries? 
\end{itemize}

\subsection{Experiment Setting}
\subsubsection{Platform} We conduct all fuzzing experiments on a server machine with 2 Intel Xeon Gold 6242R 3.10GHz 20-core CPUs and 256GB memory with the Ubuntu 20.04 operating system.
\subsubsection{Library Selection}
 We select the target libraries based on three criteria: 
\begin{itemize}
    \item The repository must have received at least 100 stars.
    \item The repository should have recent commits within the past six months.
    \item The repository must contain a minimum of five generic APIs.
\end{itemize}

We randomly select 29 libraries from crates.io and GitHub, with 10 libraries from the Trophy-case repository. Trophy-case is a repository that records the bugs found by fuzzing in the libraries. All libraries mentioned in the Trophy-case have records that have previously been discovered to contain bugs by fuzzing. Fuzzing repositories from the Trophy-case with our tool can reveal whether our tool can cover APIs that other automated generated fuzz drivers or handcraft fuzz drivers do not cover, and identify any bugs they contain. All libraries satisfy the criteria claimed above. Within these libraries, the number of APIs ranges from 16 to 1511, and the number of generic APIs ranges from 9 to 448. 

\subsubsection{fuzz driver synthesis}
We generate all fuzz drivers by our approach. As we mention in section ~\ref{sec:api-sequence-generation}, all generated fuzz drivers contain at least one generic API. We build all fuzz drivers with the command \texttt{cargo afl build}. 

\subsubsection{Fuzzing Configuration}
We determined our fuzzing configuration by referencing RULF. We fuzz all the selected libraries with afl.rs\cite{afl-rs}, which is a Rust binding for AFL++. We set two environment variables, \texttt{AFL\_EXIT\_WHEN\_DONE}=1 and \texttt{AFL\_NO\_AFFINITY}=1. \texttt{AFL\_EXIT\_WHEN\_DONE} leads to an early termination for the fuzz drivers which cannot discover new finds, while \texttt{AFL\_NO\_AFFINITY} enables us to run more fuzzing instances than the actual number of CPUs. Each fuzz driver is configured with a maximum runtime of 24 hours. 

\subsection{RQ1: Can our approach generate valid fuzz drivers to cover generic APIs?}
 Table~\ref{tab:results} shows the main results of our experiments. We answer this research question from validity and generic API coverage.

\subsubsection{Validity of generated fuzz drivers} The results show our tool reaches a high rate of validity. We generate 2 to 300 fuzz drivers for each library. The total number of generated fuzz drivers by our tool is 2566, with 103 invalid fuzz drivers. The ratio of valid fuzz drivers to invalid fuzz drivers is 0.04. After an analysis of all invalid drivers, we have discovered the main cause of invalidity is the unsound trait bounds check. As we mention in section~\ref{sec:api-graph-monomorphization}, to reach a higher generic API coverage, we implement an unsound trait bounds checker. Thus, a few invalid usages of APIs can pass the trait bounds checker and cause the generation of invalid fuzz drivers. This issue can be fixed by more engineering effort or by using the official trait bounds checker in the Rust compiler.

\subsubsection{Coverage of our approach} Table~\ref{tab:results} presents the coverage for API and generic API in each library. Our approach can achieve 0.7+ API coverage for 14 libraries and 0.6+ generic API coverage for libraries. However, it does not work well for some libraries, and the coverage of our tool varies among different libraries. Across all libraries, the API coverage spans from 0.01 to 0.96, and the generic API coverage ranges from 0.01 to 0.75. The total API coverage is 0.66, while the total generic API coverage is 0.27. To understand the culprits of the low generic API coverage, We study the library with the lowest generic API coverage (\texttt{axum-core}). We have 72 uncovered generic APIs, and we find that 68 of them are caused by using the types or traits of third-party libraries in the generic API signature in this library. Figure~\ref{fig:case-third-party} shows an example of this situation. The \texttt{axum-core} declares a trait \texttt{FromRequest} with an interface \texttt{from\_request}. However, the first argument of this interface is a third-party type \texttt{http::Request}. Since we lack the API information from \texttt{http}, we can not instantiate a \texttt{http::Request}. As a result, all APIs related to this trait are uncovered. This issue is difficult to resolve since we implement our tool based on rustdoc, which is designed for documenting a single library. It takes a lot of engineering effort to adopt our tool to extract the API information from third-party libraries. The reason for the rest of the 4 uncovered APIs is that there are no types that implement the trait required by the trait bounds of the generic API. Since we generate monomorphic API from all existing types in the library, our approach can not cover this situation.

Note that our approach is bound to outperform RULF, as our method essentially extends the API dependency graph utilized by RULF. Consequently, the covered APIs by our tool is a superset of RULF's.

\begin{figure}[htbp]
    \centering
\begin{lstlisting}[language=Rust]
use crate::body::Body;
pub type Request<T = Body> = http::Request<T>;
pub trait FromRequest<S, M = private::ViaRequest>: Sized {
    type Rejection: IntoResponse;
    async fn from_request(req: Request, state: &S) 
    -> Result<Self, Self::Rejection>;
}
\end{lstlisting}
    \caption{The generic API in \texttt{axum-core} use a third-party type from \texttt{http}. }
    \label{fig:case-third-party}
\end{figure}






\subsection{RQ2: Does our pruning algorithm improve efficiency?}

\begin{figure}[ht]
    \centering
    \includegraphics[width=1\linewidth]{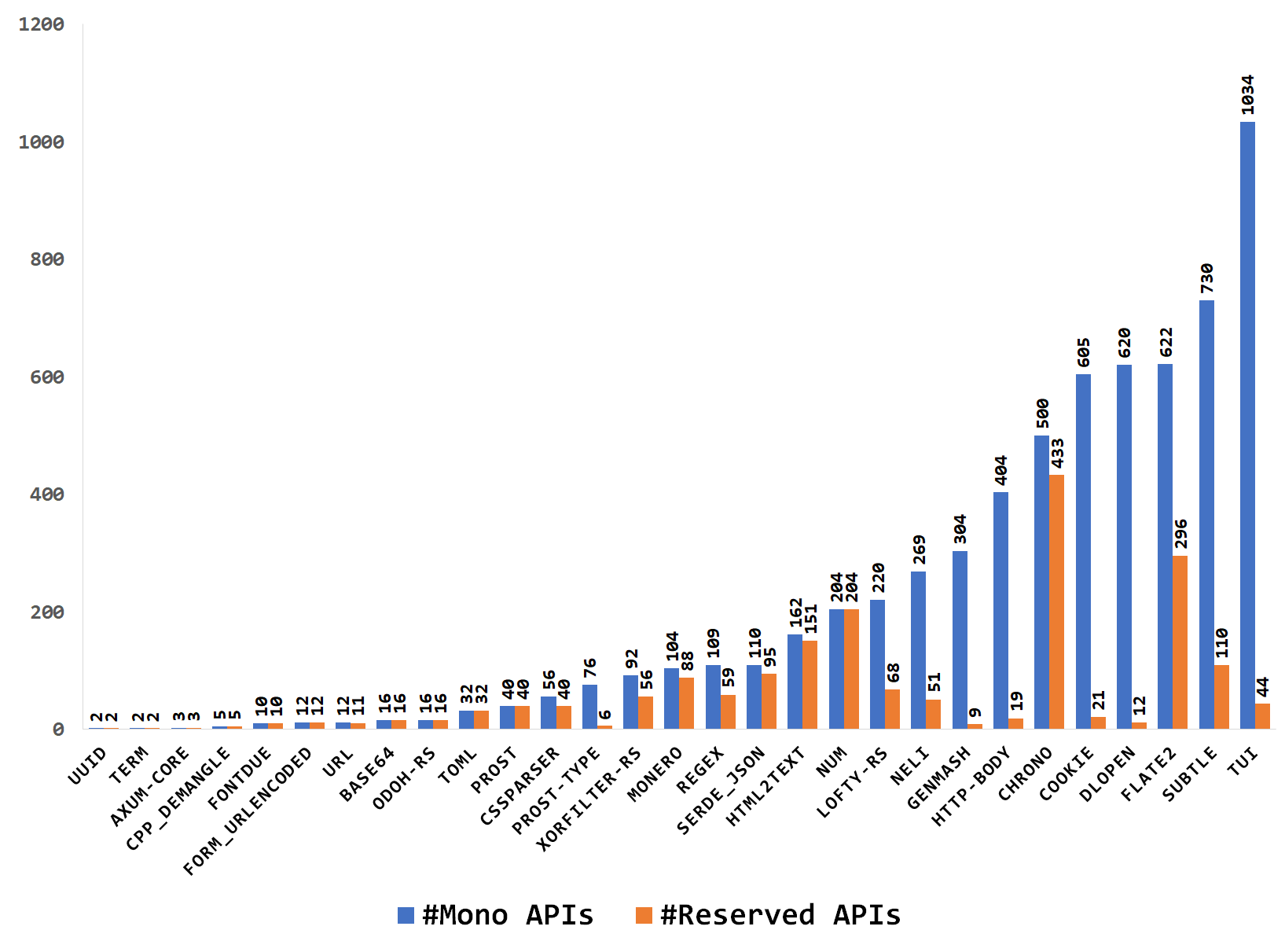}
    \caption{Comparison between \# of monomorphic APIs and \# of reserved APIs. The libraries are ordered by \# of monomorphic APIs.}
    \label{fig:efficiency}
\end{figure}

 We report the number of monomorphic APIs and reserved APIs for each library in table~\ref{tab:results}. The total number of monomorphic APIs is 6371, while the total number of reserved APIs is 1911. As a conclusion, the number of monomorphic APIs is reduced to 30\% of their original count after similarity pruning. In the best situation, the number of reserved APIs is only 2\% of the number of monomorphic APIs (in \texttt{dlopen} library). We also observe that our pruning algorithm is generally more effective in cases with a larger number of monomorphic APIs. Figure~\ref{fig:efficiency} shows the comparison between the number of monomorphic APIs and the number of reserved APIs in libraries, ordered by the number of monomorphic APIs in the library. As the number of monomorphic APIs increased, the proportion of monomorphic APIs reduced by the pruning algorithm also gradually increased. This is understandable, as a higher count of monomorphic APIs is often a result of similar behaviors.
 
\subsection{RQ3: What kinds of bugs our approach can find in the real-world libraries?}
\subsubsection{Overview}
We distinguish bugs by the crash message and stack trace. If two crashes have the same crash message, and the stack trace implies the crash happened in the same API, we regard them as the same bugs. We manually review all crash reports and filter out duplicate bugs. All bugs we found are submitted to the GitHub issue of the library. Table~\ref{tab:results} shows the bugs we found in the libraries. Experimental results indicate that our approach can identify bugs within generic APIs. After fuzzing 29 libraries, we found 23 bugs, with 10 bugs confirmed by the library maintainer before we submitted the paper. We manually investigate the culprit of all bugs, and present a taxonomy as follows.
 \begin{itemize}
     \item \textbf{Utf-8 encoding.} 2 bugs are caused by this reason.
     \item \textbf{Arithmetic overflow.} 9 bugs are caused by this reason.
     \item \textbf{Out of memory.} 6 bugs are caused by this reason.
     \item \textbf{Index out of bounds.} 2 bugs are caused by this reason.
     \item \textbf{Other.} The program crashed for miscellaneous reasons, including unreachable code, option unwrap, infinite loop, and unexpected panic. 4 bugs are caused by this reason.
 \end{itemize}

We also examine all bugs to ascertain how many bugs are related to the generic API. A bug is considered generic API-related if it is either caused by a generic API or requires the invocation of at least one generic API. Out of 23 bugs, we found 18 bugs are related to generic APIs.

\subsubsection{Case Study}
We do a further case study to understand how bugs in generic API are triggered. The figure~\ref{fig:roles} shows an example derived from our experiment, where we merge and reformat two fuzz drivers to make it more readable. These two fuzz drivers find different bugs on the same API. \texttt{parse} and \texttt{render} are both generic APIs. \texttt{parse} requires a type implemented \texttt{io::Read} trait and return a \texttt{RenderTree} instance. \texttt{render} requires a type implemented \texttt{TextDecorator} trait, and receive \texttt{RenderTree} instance from \texttt{parse}. \texttt{render} is a buggy API and contains two bugs. The first bug (line 9-10) is an arithmetic overflow bug and can be triggered when \texttt{render} receives a \texttt{RenderTree} parsed from certain data and an instance of type \texttt{RichDecorator}. The second bug (line 11-12) is an infinite loop bug and can be triggered when \texttt{render} receives a \texttt{RenderTree} parsed from certain data and an instance of type \texttt{TrivialDecorator}. Any other type implemented \texttt{TextDecorator} cannot trigger this bug. Since \texttt{RichDecorator} and \texttt{TextDecorator} implement trait \texttt{TextDecorator} independently, our approach generates and reserves these two monomorphic APIs for API \texttt{render}, has the capability to detect both bugs. However, other previous work can only detect one of these bugs since they neglect the generic API or simply consider the generic API as a non-generic API. This case study shows the fuzz drivers generated by RULF are diverse and can detect various bugs hidden in different monomorphic APIs. 

\begin{figure}[ht]
\begin{lstlisting}[language=Rust]
/* API Signatures:
fn parse(mut input: impl io::Read) -> RenderTree
fn render<D: TextDecorator>(self: RenderTree, width: usize,
decorator: D) -> RenderedText<D>  */
use html2text::render::text_renderer::RichDecorator;
use html2text::render::text_renderer::TrivialDecorator;
let data: &[u8] = get_data();
let render_tree = html2text::parse(&data);
// RichDecorator triggers an overflow bug
let _ = render_tree.render( 1, RichDecorator{}); 
// TrivialDecorator triggers an infinite loop bug
let _ = render_tree.render( 1, TrivialDecorator::new()); 
\end{lstlisting}
    \caption{Example of diverse bugs in a generic API.}
    \label{fig:roles}
\end{figure}







\section{Related Work}
 Fuzz driver synthesis can roughly be divided into two categories based on their strategies. The first kind of approach leverages the consumer code for libraries and synthesizes fuzz drivers with the call graph. The representative work includes Fudge\cite{babic2019fudge}, FuzzGen\cite{fuzzgen2020} and RUBICK\cite{zhangautomata}. Another kind of approach leverages the API specification from the source code and synthesizes fuzz drivers. The representative work includes RULF\cite{jiang2021rulf}, GraphFuzz\cite{green2022graphfuzz}. Some works, such as APICraft\cite{zhang2021apicraft}, Hopper\cite{chen2023hopper} explore the hybrid approach to leverage both API specification and consumer code. There are also other tools to test library API. Rudra\cite{bae2021rudra} is a static analyzer that analyzes Rust code and detects three types of bugs. SyRust\cite{syrust21} synthesizes the Rust program by modeling the Rust type system. However, the program synthesized by SyRust cannot serve as a fuzz driver since it cannot receive the input.

\section{Conclusion}
In this paper, we propose a novel fuzz driver synthesis approach for libraries with generic APIs. We analyze the challenges and propose a two-stage approach to generate valid fuzz drivers with generic APIs. We implement a prototype of our approach called RULF+ and evaluate our approach on 29 real-world open-source Rust libraries. Experimental results and case studies support that RULF+ can reach a high rate of valid cases and promising generic API coverage. Our pruning approach can improve the efficiency of fuzzing while still keeping the capability of finding elusive and diverse bugs in generic APIs.



\bibliographystyle{ACM-Reference-Format}
\input{main.bbl}


\appendix

\end{document}

%% file: pseudo/reachability-v2.tex
\begin{algorithm}
    \caption{reachabable monomorphic API search}
    \SetAlgoNoLine
    \label{alg:reachable-search}
    \SetKwInOut{Input}{Input}
    \SetKwInOut{Output}{Output}
    \SetKwProg{Fn}{Function}{:}{}
    \SetKwFunction{Match}{Match}
    \SetKwFunction{Merge}{Merge}
    \SetKwFunction{SatisfyTraitBound}{SatisfyTraitBounds}
    \SetKwFunction{IsGeneric}{IsGeneric}
    \SetKwFunction{IsReachable}{IsReachable}
    \SetKwFunction{UpdateMatchEdges}{UpdateMatchEdges}
    \SetKwFunction{Mono}{Mono}
    \SetKwFunction{MonoAPI}{MonoAPI}
    \SetKwFunction{Depth}{Depth}
    \SetKw{And}{and}
    \SetKw{Not}{not}
    \SetKw{Continue}{continue}
    \SetKw{Break}{break}
    \Input{
        Set of non-generic API nodes: $A$,\\
        Set of generic API nodes: $\mathcal{A}$,  \\
        Set of producer edges: $P$, \\
        Set of consumer edges: $C$, \\
        Depth limit of nested type: $maxDepth$\\
    }
    \Output{
        Set of reachable monomorphic solutions for each generic API: $MO(gapi)$
    }

    $reachableType=\emptyset$\;
    \Repeat{$reachableType$ remains unchanged}{
        \For{$api \in A$}{
            \If{\IsReachable{api}}{
                \For{$api\rightarrow returnType \in P$}{
                    $reachableType=reachableType\cup returnType$\;
                }
            }
        }
        \For{$gapi\in \mathcal{A}$}{
            $s = \{(\top, \top, \cdots)\}$\;
            \For {$p\rightarrow gapi \in C$}{
                \If{\IsGeneric{$p$}}{
                    $mo = $\{$(\bot, \bot, \cdots)$\}\;
                    \For{$ty\in reachableType$}{
                        $mo = mo\cup$ \Match{ty,p}\;
                    }
                    $s = s \cap mo$\;
                }
            }
            \For {$mo \in s$}{
                \If{\SatisfyTraitBound{$mo$, $gapi$}}{
                    $mapi=$\MonoAPI{$mo$, $gapi$}\;
                    \If {\IsReachable{mapi}}{
                        $reachableType=reachableType\cup mapi.returnType$\;
                        $MO(gapi)=MO(gapi)\cup m$\;
                    }
                }

            }
        }
    }
    \BlankLine

\end{algorithm}

%% file: pseudo/pruning.tex
\begin{algorithm}
    \caption{Similarity Pruning}
    \label{alg:similarity-pruning}
    \SetAlgoNoLine
    \SetKwInOut{Input}{Input}
    \SetKwInOut{Output}{Output}
    \SetKw{And}{and}
    \SetKw{Not}{not}
    \SetKw{Continue}{continue}
    \SetKwProg{Fn}{Function}{:}{}
    \SetKwFunction{InitDiverse}{GetMinimalSetCover}
    \SetKwFunction{AddReserve}{AddReserveAPI}
    \SetKwFunction{Pruning}{Pruning}
    \SetKwFunction{CheckReserve}{TryReserveAPI}
    \SetKwFunction{CanProduce}{CanProduce}
    \SetKwFunction{GetImpls}{Impls}
    \SetKwFunction{MonoAPI}{MonoAPI}
    \SetKwFunction{Mono}{Mono}
    \SetKwFunction{RandomSelect}{RandomSelectOne}
    \SetKwFunction{SelectMaximumUncover}{MaxUncover}
    \SetKwProg{Init}{init}{}{}
    \Input{
        Set of non-generic API nodes: $A$,\\
        Set of generic API nodes: $\mathcal{A}$,\\
        Set of producer edges: $P$,\\
        Set of reachable monomorphic solutions for each generic API: $MO(gapi)$
    }
    \Output{
        Set of reserved API $R$ \\
    }
    
    \BlankLine
    $requireType=\emptyset$, $produceType=\emptyset$\; 
    \For{$gapi\in \mathcal{A}$}{
        \InitDiverse{gapi}\;
    } 
    \Repeat{reserveType and produceType remain unchanged }{
        \For{$api\in A$}{
            \CheckReserve{api, requireType, produceType}\;

        }
        \For{$gapi\in \mathcal{A}$}{
            \For{$mo\in MO(gapi)$}{
                $mapi=$\MonoAPI{mo, gapi}\;
                \CheckReserve{mapi, requireType, produceType}\;
            }
        }
    }
    \BlankLine

    \Fn{\InitDiverse{gapi}}{
        $covered=\emptyset$, $selected=\emptyset$\;
        \Repeat{no monomorphism is selected}{
            $selectMono=$\SelectMaximumUncover{$M_{gapi}-selected$, covered, gapi}\;
            $selected=selected\cup {selectMono}$\;
            $covered=covered\cup Impls(gapi,selectMono)$
        }
        
        \If{$selected\subseteq \emptyset$}{
            $selected=selected\cup$\RandomSelect{$M_{gapi}$}\;
        }
        \For{$m\in selected$}{
            \AddReserve{\MonoAPI{m,gapi}}\;
        }
    }

    \Fn{\CheckReserve{api}}{
        \lIf{$api\notin R$}{
            \Return
        }
        $success=False$\;
        \For{$ty\in requireType$}{
            \If{$ty\notin produceType$}{
                \If {$api\rightarrow ty \in P$}{
                    $produceType=produceType\cup ty$\;
                    $success=True$\;
                }
            }
        }
        \lIf{$success$}{
            \AddReserve{api}
        }
    }
    
    \Fn{\AddReserve{api}}{
        $R=R\cup api$\;
        \For{$p \in api.args$}{
            $reserveType=reserveType\cup p.type$\;
        }
    }

\end{algorithm}

%% file: tables/transformation.tex
\begin{table}[ht]
  \centering
  \caption{Predefined transformation rules.}
  \label{tab:transformation}
  \begin{tabular}{lll}
    \toprule
     Borrow Rules  &  Pointer Rules        & Wrapper Rules \\
    \midrule
     T$\to$\&T     &  T$\to$ $\ast$const T & Result<T ,E> $\to$T \\
     T$\to$\&mut T &  T$\to$ $\ast$mut T   & Option<T> $\to$T \\
     \bottomrule
\end{tabular}
\end{table}

%% file: tables/result.tex
\begin{table*}[ht]
\centering
\caption{Experiment results of 29 libraries. The first 18 are from GitHub, the last 11 are from Trophy-case. "\#Mono APIs" represent the number of monomorphic APIs obtained by reachable monomorphic API search.}
\begin{tabular}{c|c|c|c|c|c|c|c|c|c|c}
\toprule
Library &
  Version &
  \#APIs &
  \begin{tabular}[c]{@{}c@{}}\#Generic\\ APIs\end{tabular} &
  \begin{tabular}[c]{@{}c@{}}API\\ Coverage\end{tabular} &
  \begin{tabular}[c]{@{}c@{}}Generic API \\ Coverage\end{tabular} &
  \begin{tabular}[c]{@{}c@{}}\#Mono \\ APIs\end{tabular} &
  \begin{tabular}[c]{@{}c@{}}\#Reserved\\ APIs\end{tabular} &
  \begin{tabular}[c]{@{}c@{}}Fuzz targets\\ (invalid)\end{tabular} &
  Crashes &
  Bugs \\
  \midrule
serde\_json      & 1.0.103 & 586  & 317 & 0.49 & 0.10 & 110  & 95  & 300 (33)& 103        & 0 \\
base64           & 0.21.4  & 40   & 22  & 0.60 & 0.64 & 16   & 16  & 17 (4)  & 0          & 0 \\
uuid             & 1.4.1   & 133  & 9   & 0.94 & 0.22 & 2    & 2   & 7       & 1          & 0 \\
form\_urlencoded & 1.2.0   & 16   & 9   & 0.81 & 0.67 & 12   & 12  & 10      & 4          & 1 \\
flate2           & 1.0.28  & 274  & 228 & 0.54 & 0.50 & 622  & 296 & 147 (3) & 66         & 0 \\
http-body        & 0.4.5   & 156  & 143 & 0.17 & 0.09 & 404  & 19  & 6 (4)   & 0          & 0 \\
subtle           & 2.5.0   & 107  & 19  & 0.96 & 0.79 & 730  & 110 & 300     & 17         & 0 \\
prost-type       & 0.12.1  & 453  & 17  & 0.95 & 0.12 & 76   & 6   & 23      & 0          & 0 \\
axum-core        & 0.3.4   & 95   & 73  & 0.01 & 0.01 & 3    & 3   & 3 (2)   & 0          & 0 \\
monero           & 0.19.0  & 469  & 148 & 0.66 & 0.32 & 104  & 88  & 105     & 554        & 3 \\
dlopen           & 0.1.8   & 55   & 48  & 0.29 & 0.25 & 620  & 12  & 2       & 0          & 0 \\
lofty-rs         & 0.17.1  & 1186 & 144 & 0.73 & 0.15 & 220  & 68  & 27 (2)  & 30         & 2 \\
html2text        & 0.6.0   & 145  & 78  & 0.72 & 0.67 & 162  & 151 & 182 (6) & 8515       & 5 \\
neli             & 0.7.0   & 1511 & 448 & 0.61 & 0.04 & 269  & 51  & 10      & 10         & 4 \\
odoh-rs          & 1.0.2   & 39   & 15  & 0.74 & 0.80 & 16   & 16  & 21      & 6          & 2 \\
xorfilter-rs     & 0.5.1   & 48   & 35  & 0.85 & 0.80 & 92   & 56  & 56      & 70         & 3 \\
genmesh          & 0.6.0   & 31   & 30  & 0.13 & 0.13 & 304  & 9   & 6       & 0          & 0 \\
term             & 0.7.0   & 42   & 19  & 0.26 & 0.11 & 2    & 2   & 4       & 0          & 0 \\
\midrule
chrono           & 0.4.31  & 639  & 202 & 0.88 & 0.73 & 500  & 433 & 300     & 347        & 0 \\
cookie           & 0.18.0  & 116  & 34  & 0.78 & 0.44 & 605  & 21  & 22 (13) & 0          & 0 \\
cssparser        & 0.33.0  & 148  & 49  & 0.72 & 0.41 & 56   & 40  & 201 (2) & 0          & 0 \\
regex            & 1.8.3   & 207  & 16  & 0.93 & 0.69 & 109  & 59  & 57      & 258        & 2 \\
tui              & 0.19.0  & 366  & 71  & 0.82 & 0.42 & 1034 & 44  & 45 (22) & 47         & 1 \\
url              & 2.4.1   & 113  & 12  & 0.90 & 0.75 & 12   & 11  & 67 (2)  & 0          & 0 \\
toml             & 0.8.2   & 239  & 161 & 0.38 & 0.14 & 32   & 32  & 300 (8) & 22         & 0 \\
prost            & 0.12.1  & 282  & 183 & 0.37 & 0.13 & 40   & 40  & 127     & 0          & 0 \\
num              & 0.4.1   & 56   & 56  & 0.36 & 0.36 & 204  & 204 & 204     & 265        & 1 \\
fontdue          & 0.8.0   & 54   & 12  & 0.39 & 0.17 & 10   & 10  & 12 (2)  & 0          & 0 \\
cpp\_demangle    & 0.4.3   & 104  & 11  & 0.17 & 0.27 & 5    & 5   & 5       & 0          & 0 \\
\midrule
\textbf{Total}     & N/A     & 7710 & 2609  & 0.66 & 0.27 & 6371    & 1911   & 2566 (103)       & 10319          & 23 \\
\bottomrule
\end{tabular}

\label{tab:results}
\end{table*}